\newif\ifanonymousversion
\anonymousversionfalse

\newif\ifarxivversion
\arxivversiontrue

\ifanonymousversion
  \documentclass[sigconf,anonymous]{acmart}
  \setcopyright{none}
  \renewcommand\footnotetextcopyrightpermission[1]{}
\else
  \ifarxivversion
    \documentclass[sigconf]{acmart}
      \setcopyright{none}
      \renewcommand\footnotetextcopyrightpermission[1]{}
  \else
    \documentclass[sigconf]{acmart}
  \fi
\fi

\usepackage{graphicx}
\usepackage{xcolor}
\usepackage{soul}

\begin{document}

\pagestyle{plain}

\title{Security Evaluation of Thermal Covert-channels on SmartSSDs}

\ifanonymousversion

\author{Anonymous Submission \vspace{2cm}}

\else

\author{Theodoros Trochatos}
\affiliation{%
  \institution{Yale University}
  \city{New Haven}
  \state{CT}
  \country{USA}
}
\email{theodoros.trochatos@yale.edu}

\author{Anthony Etim}
\affiliation{%
  \institution{Yale University}
  \city{New Haven}
  \state{CT}
  \country{USA}
}
\email{anthony.etim@yale.edu}

\author{Jakub Szefer}
\affiliation{%
  \institution{Yale University}
  \city{New Haven}
  \state{CT}
  \country{USA}
}
\email{jakub.szefer@yale.edu}

\renewcommand{\shortauthors}{Theodoros Trochatos, Anthony Etim, and Jakub Szefer}

\fi

\date{}
\maketitle

\section*{Abstract}

Continued expansion of cloud computing offerings now includes SmartSSDs. A SmartSSD is a solid-state disk (SSD) augmented with an FPGA. Through public cloud providers, it is now possible to rent on-demand virtual machines enabled with SmartSSDs. Because of the FPGA component of the SmartSSD, cloud users who access the SmartSSD can instantiate custom circuits within the FPGA. This includes possibly malicious circuits for measurement of power and temperature. Normally, cloud users have no remote access to power and temperature data, but with SmartSSDs they could abuse the FPGA component to learn this information. This paper shows for the first time that heat generated by a cloud user accessing the SSD component of the SmartSSD and the resulting temperature increase, can be measured by a different cloud user accessing the FPGA component of the same SmartSSD by using the ring oscillators circuits to measure temperature. The thermal state remains elevated for a few minutes after the SSD is heated up and can be measured from the FPGA side by a subsequent user for up to a few minutes after the SSD heating is done. Further, in a future multi-tenant SmartSSD setting, the thermal changes can be measured in parallel if one user controls the SSD and the other the FPGA. Based on this temporal thermal state of the SmartSSD, a novel thermal communication channel is demonstrated for the first time.

\section{Introduction}
\label{sec_introduction}

The business model of cloud computing focuses on temporal sharing of the hardware between users. When one user is not using the hardware, it can be assigned to other users. Cloud providers such as Amazon now charge by the minute or even by the second for certain virtual machine instance types~\cite{awsnewsblog}. In addition to temporal sharing, there is also the possibility of spatial sharing. A particular piece of hardware (such as CPU or FPGA) can be assigned to multiple users at the same time. This is common for CPUs, but also has been explored for cloud-based FPGAs~\cite{khawaja2018sharing}.

Cloud computing has originally focused on using CPUs. Later GPUs were added and FPGAs that nowadays can be rented for remote access from cloud. Until recently, the main FPGA-enabled devices available from the various cloud providers were FPGA accelerator cards, consisting of an FPGA chip and a few dedicated DRAM modules on each FPGA accelerator card. Recently, however, a new offering has been introduced: the SmartSSD~\cite{smartssd}. SmartSSD is a solid-state disk (SSD) augmented with an FPGA. The disk and FPGA share a PCIe connection to the host computer and are enclosed in a single package. The purpose of the FPGA is to enable computation on the data stored on the disk, without use of the main host computer. Through public cloud providers such as VMAccel~\cite{vmaccel} it is now possible to rent SmartSSDs on-demand. Following the recent introduction of the SmartSSD into the cloud computing environment, this paper is the first paper to explore the security of SmartSSDs in the cloud.

In particular, this work presents a new thermal covert communication channels that leverages the thermal state of the SmartSSDs. The thermal channel is shown to be extremely easy to establish. To transmit information, the sender can either stress the SSD by accessing large amounts of data (to generate heat and send $1$) or do nothing (to keep the temperature low and send $0$). As this paper shows for the first time, the thermal changes can be observed on the FPGA component of the SmartSSD, because of the heat transfer within the SmartSSD package. Because the SSD and FPGA are in the same enclosure, the thermal changes due to activity of the SSD affect the temperature of the FPGA chip. This paper evaluates the thermal behavior of the SmartSSDs both in a remote server in our university and on a public cloud provider that offers SmartSSD enabled virtual machines.

This paper shows for the first time that a thermal covert channel can be establised by cloud user accessing the SmartSSD. We show that a user can heat up the SSD component of the SmartSSD and the resulting temperature increase can be measured by a different cloud user accessing the FPGA component of the same SmartSSD. Typically, cloud users do not have access to thermal information, as this is blocked by the cloud provider. However, by instantiating ring oscillators in the FPGA, cloud users can bypass typical cloud protections and remotely measure the temperature (in this case of the SmartSSD).

The covert channel based upon the temporal thermal state can be used without use of error correcting codes (ECC) for the transmission of information between two users who gain sequential access to the same SmartSSD. This already includes the effects of data center cooling system, which constantly cools the servers and the disks and it cannot be controlled by the attacker as the SmartSSD disks are accessed remotely by the cloud users. Furthermore, due to the abundance of cloud computing resources, multiple disks can be easily rented in parallel to increase the bandwidth of the covert transmission in proportion to the number of disks uses. In addition, the thermal changes of the SSD can be observed on the FPGA in parallel as the SSD is heating up. Thus, if SSD and FPGA, or part of FPGA, are allocated to different users in a potential multi-tenant setting, the FPGA can be used to spy on the SSD activity.

\subsection{Contributions}

This paper makes a number of new contributions:
\begin{itemize}

\item We introduce the first security evaluation of the new SmartSSDs in a cloud setting.
\item We present a novel thermal covert communication channel leveraging FPGAs inside the SmartSSDs, the channel is based on  using ring oscillators that can be instantiated by malicious cloud users.
\item We evaluate the SmartSSDs' thermal behavior and properties on a university server and on a public cloud provider.
\item We present the first covert communication channel in SmartSSDs, approaching $100$\% accuracy and in which the bandwidth can be easily scaled by using multiple SmartSSDs in parallel.

\end{itemize}

\section{Background}
\label{background}

Cloud computing is now an established computing paradigm. However, there are constantly new devices being made available for remote access. CPUs and GPUs have been available for many years, but more recently FPGAs (Field Programmable Gate Arrays) are also available, and now SmartSSDs. One of the first public cloud providers offering FPGA-accelerated virtual machine instances to users, since around 2016, was Amazon Web Services (AWS)~\cite{aws_developer_2016}.
One of the newest public cloud provider offering different types of devices is VMAccel~\cite{vmaccel}. VMAccel specializes in providing FPGA as a Service (FaaS), where users can easily deploy existing FPGA code or develop new bitstreams in their pre-configured development environments, for example. In addition to numerous FPGAs, VMAccel enables users to access the SmartSSDs.

\subsection{SmartSSDs}

\begin{figure}[t]
     \centering
         \centering
         \includegraphics[width=0.395
         \textwidth]{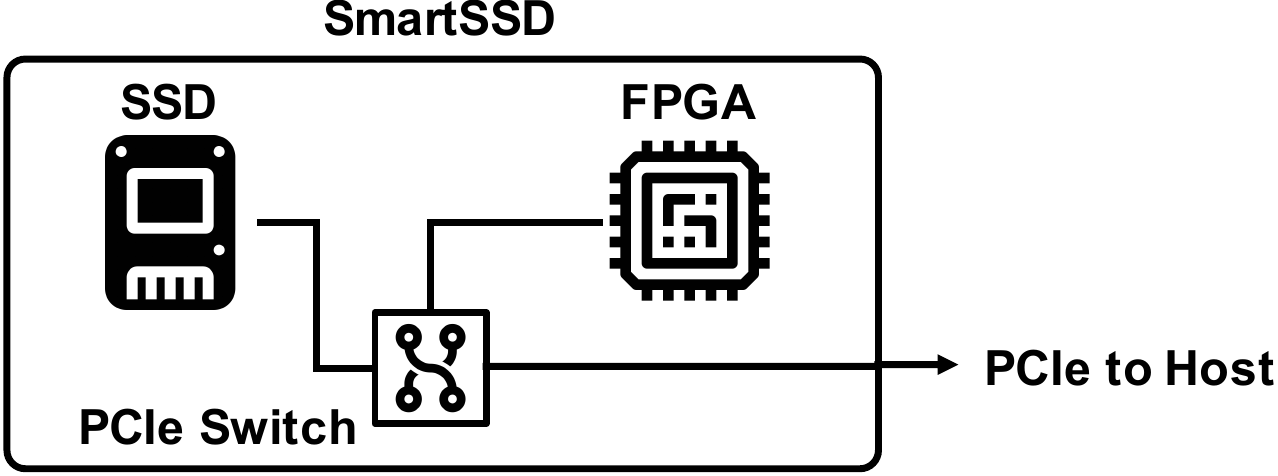}
         \caption{Block diagram of a SmartSSD.}
         \label{fig_high_level_diagram}
\end{figure}

A block diagram of a SmartSSD is shown in Figure~\ref{fig_high_level_diagram}. The SmartSSD developed by Samsung~\cite{smartssd} contains an SSD for data storage, as well as a Xilinx Kintex™ Ultrascale+ KU15P FPGA for data processing. The two components have access to the PCIe bus, which is also used to connect to the host computer. Importantly, the two devices are contained in a same package and share the PCIe and power supply (from the PCIe). By the nature of the packaging, as we show, they are also mutually affected by thermal changes - that is to say, if the SSD is heated up, this affects the temperature of the FPGA.

\subsection{Cloud-based Access to SmartSSDs}

Similar to other cloud-based computing resources, SmartSSDs are now offered as a cloud-based service, where users can get pay-as-you-go access to SmartSSDs. A typical cloud computing model is a ``single-tenant'' model where a user gets access to the whole device and when they are done the device is allocated to another user. In a ``multi-tenant'' model, multiple users may be assigned to the same device at the same time. While not yet available in the context of SmartSSDs, multi-tenant setting could include one user accessing the SSD, while another accesses the FPGA. An alternative model would be multiple users share the SSD and the FPGA at the same time. Multi-tenant FPGAs (outside SmartSSD setting) have been actively explored in academia~\cite{8988652}.

\subsection{Thermal Measurements with ROs}

Ring Oscillators (ROs) are circuits which can be instantiated inside FPGAs and can be used to measure temperature or voltage changes~\cite{10.5555/647924.738726}. By using ROs, malicious users can bypass security protections that may try to limit access to thermal or voltage data. Because SmartSSDs contain an FPGA, as this work shows for the first time, ROs can be instantiated inside the FPGA of the SmartSSD to measure thermal changes.

A ring oscillator thermal sensor in an FPGA \cite{10.5555/647924.738726} can be built by using an odd number of inverters which are connected in a loop.
To bypass any Design Rule Checks (DRCs) imposed by cloud providers, an additional Flip-Flop or Latch can be inserted in the loop for more obfuscation. In our design, we used ROs with LUTs. The RO sensor works by counting the number of oscillations of the loop, compared to a reference counter generated by a crystal oscillator. The delay through the inverters and wires of the RO depends on the temperature, while the crystal oscillator used for the reference counter is not significantly affected by temperature~\cite{7954123}.

The RO sensors can be realized as an RTL kernel inside an HLS (High-Level Synthesis) based design, following one of the RTL kernel tutorials~\footnote{\url{https://github.com/Xilinx/Vitis\_Accel\_Examples}}. LUT-based ring oscillators with $3$ stages were used for the sensor~\cite{latchedRO}. The directive {\tt ALLOW\_COMBINATORIAL\_LOOPS = "TRUE"} was used to ensure combinatorial loops were allowed. Because of the directives, the XDC configuration files did not have to be modified. The tools did not block this type of ring oscillator, but other ring oscillators based on latches~\cite{latchedRO} or flip-flops~\cite{8988652} could be used if the LUT based are blocked.

\section{Threat Model}
\label{threat_model}

This work assumes a typical cloud-computing setting where users are allocated to hardware, they pay for and when a user is done using the hardware, it is allocated to another user. We assume a sender is a user who aims to covertly communicate sensitive data to a receiver and that they both use the SmartSSD to perform the covert communication. We assume that in this cloud-computing setting, the sender and receiver are able to be allocated to the same SmartSSD and that sender and receiver can reliably be scheduled one after the other on the same SmartSSD. Since SmartSSDs contain an FPGA component, existing research on cloud-based FPGA fingerprinting can be used to identify an FPGA (and thus a SmartSSD). The fingerprints can be used by sender and receiver to establish whether they have found a common SmartSSD.

We assume the cloud provider may block access to thermal sensors of the SmartSSD. However, because of the access to the FPGA component of the SmartSSD, we assume the users can instantiate any circuit they wish in the FPGA, including an RO sensor to spy on thermal and voltage changes.

\section{Covert Channel Design}
\label{sec_covert_channel_design}

\begin{figure*}[t]
     \centering
         \centering
         \includegraphics[width=0.95\textwidth]{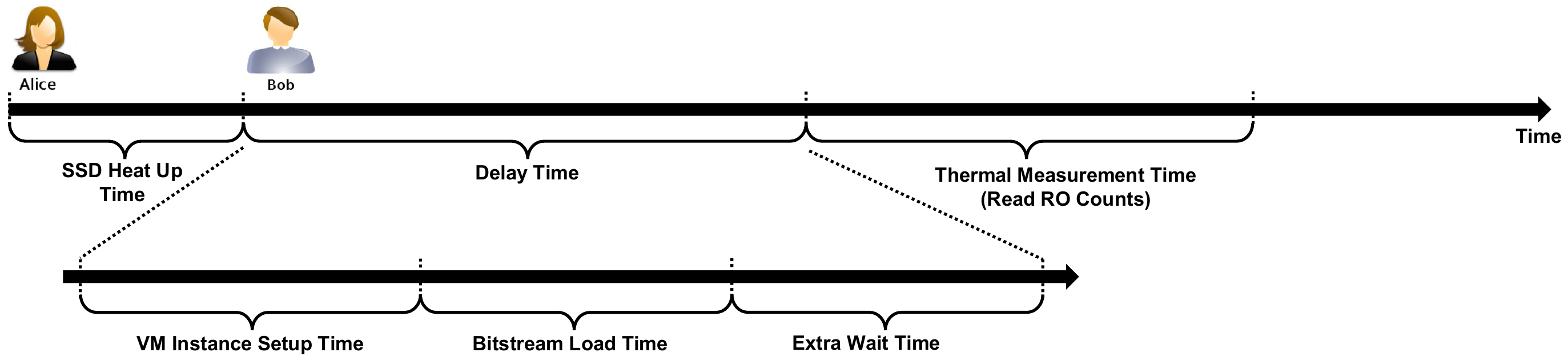}
         \caption{\small Timeline demonstrating the new covert channel between Alice and Bob, who share access to the same cloud-based SmartSSD. For simplicity, we include the time required for the users to switch in the VM Instance Setup Time.
         The time in the timeline is not shown to scale.}
         \label{fig:timeline}
     \hfill
\end{figure*}

Figure~\ref{fig:timeline} shows the design of the covert channel for two users targeting to covertly communicate information via the SmartSSD. First, Alice starts her virtual machine, obtains and heats up the SmartSSD by running the {\tt Flexible IO (FIO)} SSD stress test, discussed later. Next, Alice terminates her instance. Following that, Bob starts up a new instance with access to the same SmartSSD. After the VM instance is started, the FPGA bitstream with the RO sensors is loaded. Finally, thermal measurements are taken to learn the thermal state of the SSD and thus the information is transmitted by Alice. In the evaluation, there may also be extra waiting time, as it is shown in the figure, to account for different delays between Alice and Bob.

In this covert channel, Alice transmits one bit of data by either heating up the SSD (transmits $1$) or staying idle (transmits $0$). As we evaluate later, multiple SmartSSDs can be used in parallel to increase the bandwidth proportionally to the number of SmartSSDs~used.

\section{Experimental Setup}
\label{sec_setup}

This work evaluated SmartSSDs both on a university server and on a public cloud computing platform from which access to SmartSSDs can be rented.%
\footnote{The name of the public cloud provider used is withheld from the paper.}

\subsection{University Remote Server}

Our university server setup consists of a Linux Ubuntu server equipped with one SmartSSD disk attached to the PCIe port. The server was located in a shared server rack, emulating a simple server room or data center setup. 
Xilinx Vitis tools version $2021.1$ was used to compile the FPGA designs loaded onto the FPGA located inside the SmartSSD. Xilinx XRT version $2.11.634$ and shell version {\tt xilinx\_u2\_gen3x4\_xdma\_gc\_base\_2} were used.

\subsection{Public Cloud Server}

For the public cloud setup, we rented access to a virtual machine enabled with SmartSSDs. With the cloud provider we selected, we rented virtual machines with either one or two SmartSSDs. The main difference between the public cloud server and remote university server is the more professional cooling infrastructure, which causes SmartSSDs to operate at lower temperatures than in our remote university server.

\subsection{SSD Stress Tests Used}

\begin{table}[!t]
\centering
\caption{\small Parameters of the {\tt FIO} stress tests.}
\label{table_parameters_fio}
\footnotesize
\begin{tabular}{|c|c|c|c|c|}
\cline{1-2}\cline{4-5}
\textbf{Parameter} & \textbf{Values Tested}        & & \textbf{Parameter} & \textbf{Values Tested} \\ \cline{1-2}\cline{4-5} 
{\tt numjobs} & {1, 2, 4, 8, 16, 32, 64}           & & {\tt bs} (KB) & {64} \\ \cline{1-2}\cline{4-5}
{\tt size} (GB) & {1, 2, 4, 8, 16, 32, 64, 128}    & & {\tt iodepth} & {16} \\ \cline{1-2}\cline{4-5}
{\tt runtime} (secs) & {60, 70, 80, 120, 240, 300} & & { \tt time\_based} & {true} \\ \cline{1-2}\cline{4-5}
{\tt ioengine} & {posixaio}                        & & { \tt end\_fsync} & {true} \\ \cline{1-2}\cline{4-5}
{\tt rw} & {randwrite} \\ \cline{1-2}
\end{tabular}

\end{table}

In order to stress the SSD inside the SmartSSD, we used the {\tt Flexible IO (FIO)} tester as the stress test. The variable parameters of the stress test are: {\tt numjobs, size, runtime, ioengine, rw, bs, iodepth, time\_based} and {\tt end\_fsync}\footnote{FIO's Documentation can be found here: \url{https://fio.readthedocs.io/en/latest/}}.
Table~\ref{table_parameters_fio} shows the different parameter values that were used for stressing the disk.

Typically, we executed the stress tests on the SmartSSD disk on the public cloud provider for $60$, $120$, $240$ and $300$ seconds. For our university server, we opted for lower runtimes to prevent disk damage, since the baseline temperature for our disk is already high, compared to the cloud provider's. We set runtime for $60$, $70$ and $80$ seconds for the university server.

\subsection{Thermal Measurements Methods Used}

To measure the temperature, we used three methods:

\begin{enumerate}
    \item The {\tt nvme} utility was used to read the SSD temperature -- this would not be available to an attacker but is used to get ground truth information.
    \item The {\tt xbutil} utility also reports the FPGA temperature from a single on-chip thermal diode -- this would also not be available to an attacker but is used to get ground truth information.
    \item We developed an FPGA module that used the RTL kernel to instantiate an RO. This module can be used to estimate the temperature without need for access to any of the thermal diodes on the SSD or FPGA~chips.
\end{enumerate}

\section{ SmartSSD Characterization}
\label{sec_results}

In this section, we present our experimental results and evaluation of the different behaviors of the SmartSSDs.

\subsection{Finding Optimal SSD Heating Parameters}

We tested different values for the {\tt size} and {\tt numjobs} to understand which configuration increased the SSD temperature the most. The goal is to help us understand how a malicious user could best raise the temperature of the SSD, and by extension, the adjacent FPGA chip in the SmartSSD package.

Due to limited space, detailed data for different configurations cannot be reported here, but we observed that {\tt numjobs} $\approx 4$ and {\tt size} $\approx 8$ (GB) cause the disk to increase the most in temperature when the {\tt runtime} $=60$ seconds.
Having selected the {\tt numjobs} $=4$, we evaluated how the duration of the stress test affects the temperature for different runtimes and sizes. 
We observed that we achieve the highest temperature for {\tt size} $\approx 8$ (GB) again for both SSD and FPGA on public cloud and university remote servers. As the {\tt runtime} increases, both SSD and FPGA temperatures increase. Thus, adjusting runtime can be used to raise SSD to different~temperatures. 

We observed that the university remote server has higher baseline temperature than the public cloud server. The reason for this is that the public cloud server likely has a more capable cooling system than we have in our university remote server.
We observed that university remote server achieves almost the same relative temperature increase with only about half of the runtime that public cloud server needs. Thus, the period of heating the SSD is faster on university server.

\subsection{Duration of SSD Heating Effect}
\label{sec_stress_ref_temp_sequential}

\begin{figure}[t]
     \centering
         \centering
         \includegraphics[height=4.5cm]{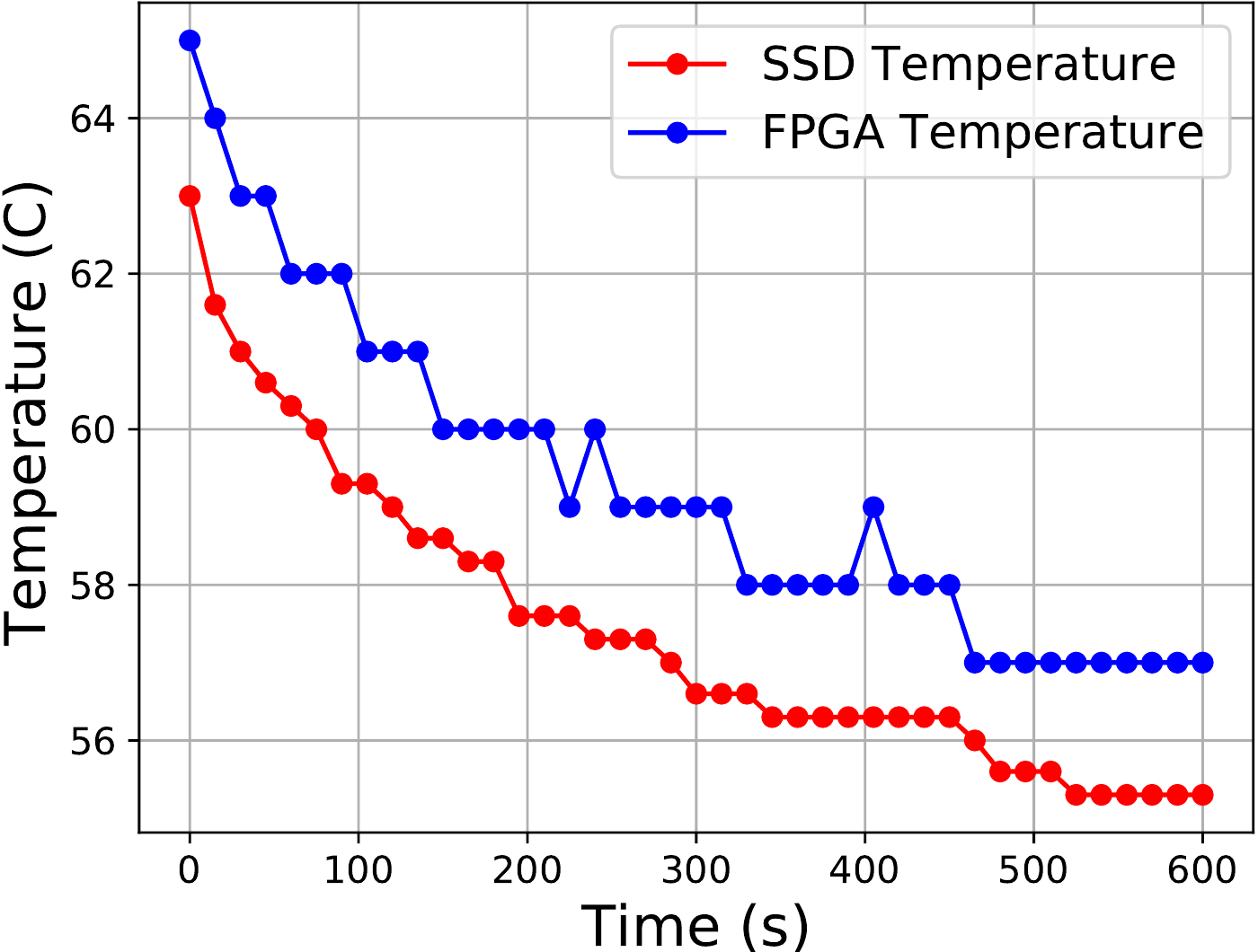}
         \caption{\small Temperature as a function of time after the stress test has stopped for the public cloud server, maximum measured time was $10$~minutes.}
         \label{fig:Total Measured Time = 10 mins vmaccel}
     \hfill
\end{figure}

\begin{figure}[t]
     \centering
         \centering
         \includegraphics[height=4.5cm]{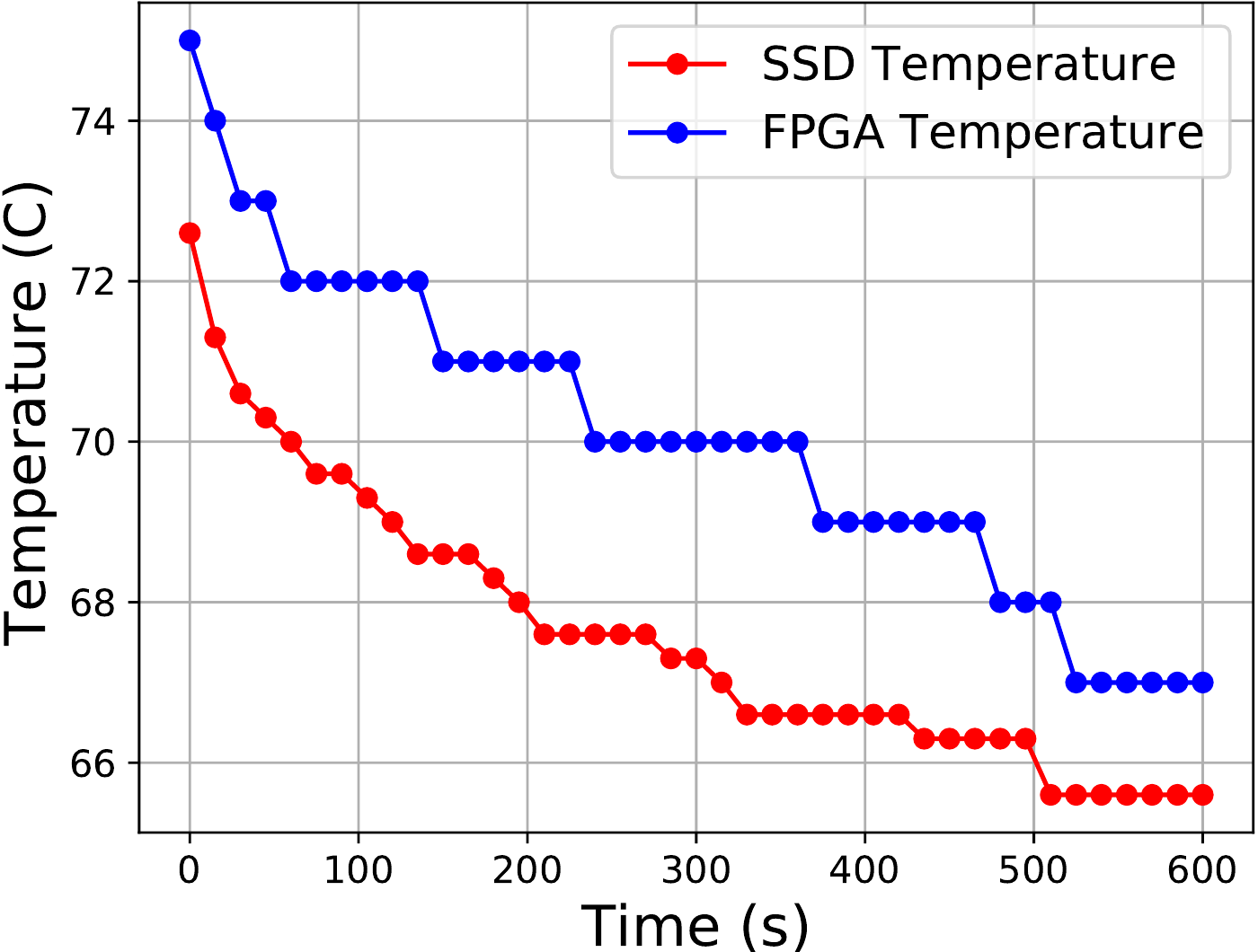}
         \caption{\small Temperature as a function of time after the stress test has stopped for the university remote server, maximum measured time was $10$~minutes.}
         \label{fig:Total Measure Time = 10 mins_cloudfpga}
     \hfill
\end{figure}

For different initial SmartSSD temperatures, we measured the temperature at different times after the stress test has concluded. We stressed the disk to reach some initial $T_i$ temperature by running several {\tt FIO} tests sequentially. After this, we performed the measurements at $15$ second intervals, while the SSD temperature cools down. We used the {\tt nvme} and {\tt xbutil} utilities to measure the SSD and FPGA temperatures, respectively. The objective of the tests is to analyze how long the SmartSSD (that means both the SSD and FPGA inside it) can retain thermal state, which later is used for the covert communication.

Figure~\ref{fig:Total Measured Time = 10 mins vmaccel} 
and Figure~\ref{fig:Total Measure Time = 10 mins_cloudfpga} show 
 how the SSD and FPGA temperatures change after performing the stress tests over a total time of $10$ minutes on the public cloud server and the university server, respectively. It can be observed that on the public cloud server, both SSD and FPGA temperatures need at least $10$ minutes to return to the baseline temperatures, if they are sufficiently heated. The same effect can be observed on the university server, despite different cooling and different baseline temperatures. 

This $10$ minute time-window can give potential malicious users sufficient time to perform the new thermal temporal covert communication, since the thermal state of the SmartSSD persists for some time. After SmartSSD is heated up, there is sufficient time for another user to read its thermal state.

\subsection{Measuring SSD Heating with ROs}
\label{sec_sequential_experiments}

\begin{figure}[t]
     \centering
         \centering
         \includegraphics[height=4.5cm]{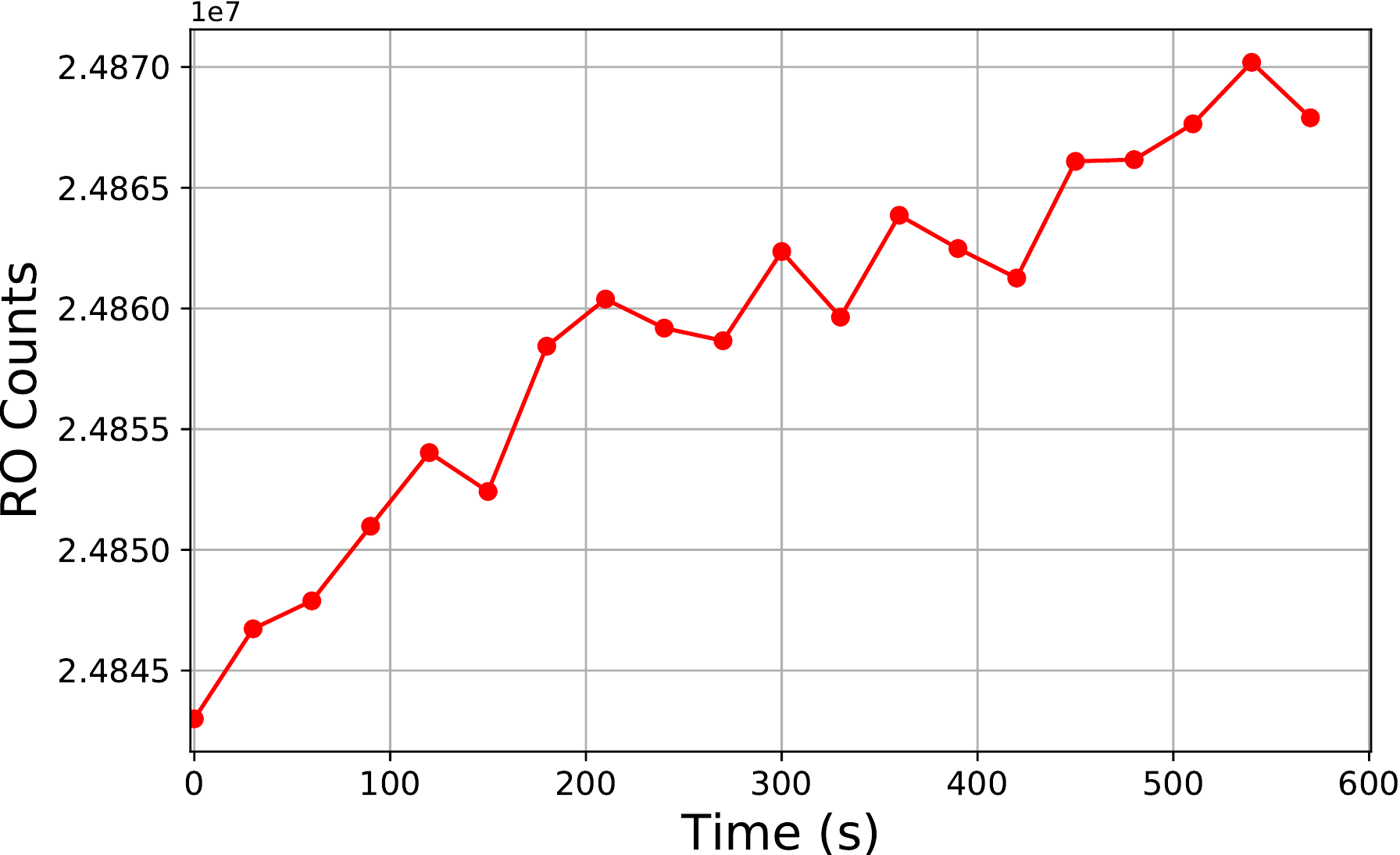}
         \caption{\small University remote server RO counts after stress tests, showing RO count increase as the SSD temperature cools off back to baseline.}
         \label{fig:Cloud FPGA Ro counts after stress test}
     \hfill
\end{figure}

\begin{figure}[t]
     \centering
         \centering
         \includegraphics[height=4.5cm]{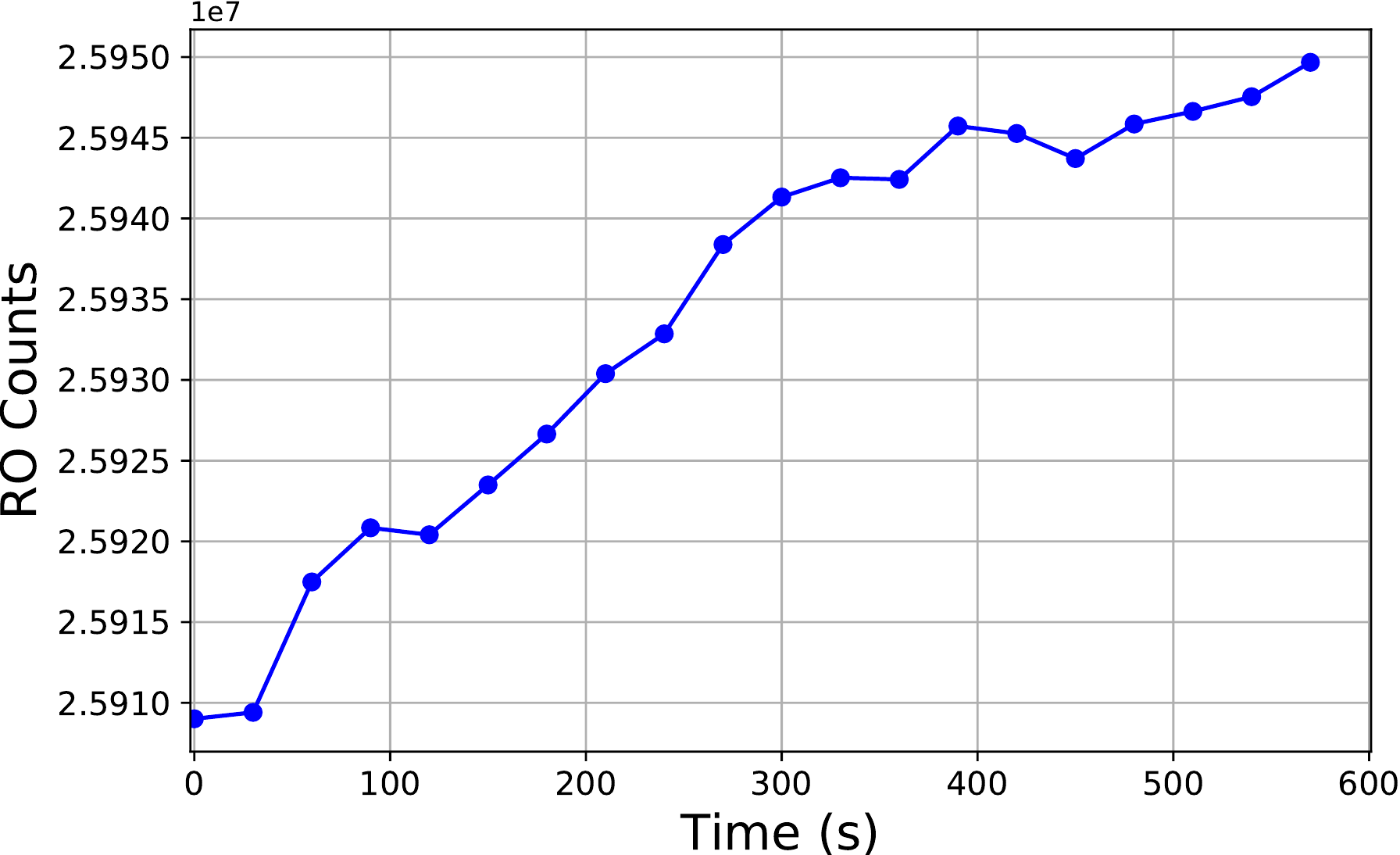}
         \caption{\small Public cloud server RO counts after stress tests, showing RO count increase as the SSD temperature cools off back to~baseline.}
         \label{fig:vmaccel Ro counts after stress test}
     \hfill
\end{figure}

In this section, we demonstrate that an attacker could measure the SmartSSD temperature using ROs, instead of doing temperature measurements using the SSD disk and FPGA utilities, which could be easily blocked by the cloud provider. To demonstrate this, we run the RO measurements on the FPGA part of the SmartSSD, while the disk cools down, after the SSD has been heated using the {\tt FIO} stress test. We stressed the disk to reach target maximum T\textsubscript{max} temperature and then we perform the measurements at $30$ seconds intervals, while the disk temperature cools down.

Figure~\ref{fig:Cloud FPGA Ro counts after stress test} shows the RO measurements for $10$ minutes after the SSD was heated up using stress test on our university remote server. At each interval, $150$ RO counts measurements were taken. Initially, the SSD disk was heated to a temperature of $73C$, which is almost $10C$ higher than the baseline SSD temperature of our university remote server. We expect for higher SSD temperature to observe lower RO counts. That is, as the disk cools down, RO counts should increase since the temperature drops, as it is clearly shown in Figure~\ref{fig:Cloud FPGA Ro counts after stress test}.

We observe similar behavior for the public cloud server, as it is shown in Figure~\ref{fig:vmaccel Ro counts after stress test}. With lower baseline temperature, the RO counts tend to be higher on the public cloud server, but also with better cooling system, more stress tests are needed to raise the temperature to $10C$ above baseline. With longer stress test, however, similar patterns in temperature are observed and the disk needs a few minutes to return to baseline temperature.

\section{SmartSSD Covert Channel Results}
\label{thermal temporal attack}

\begin{figure}[t]
     \centering
         \centering
         \includegraphics[height=4.5cm]{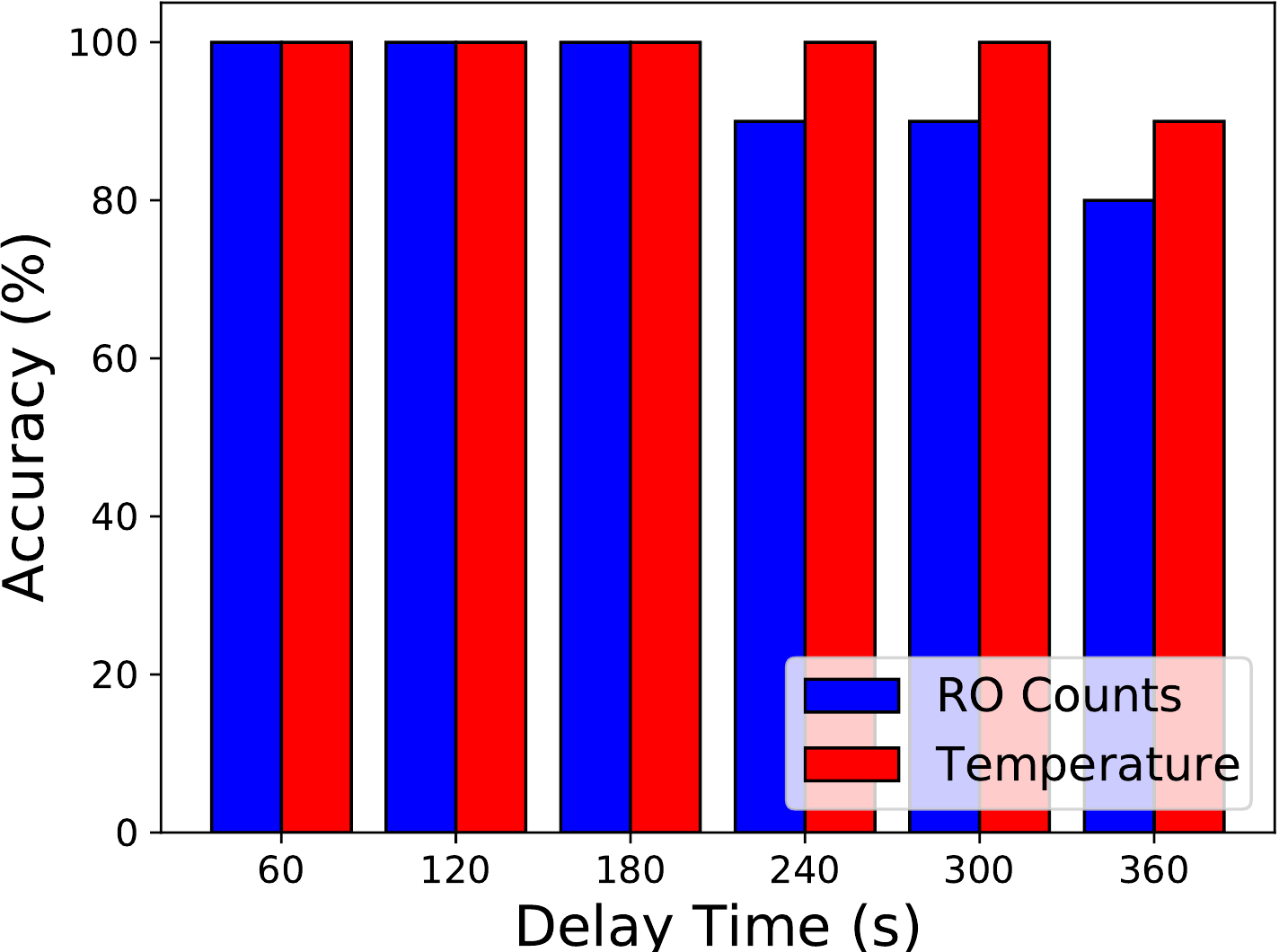}
         \caption{University remote server covert channel transmission test accuracy with different delay times.}
         \label{fig:University remote server covert channel transmission test accuracy}
     \hfill
\end{figure}

\begin{figure}[t]
     \centering
         \centering
         \includegraphics[height=4.5cm]{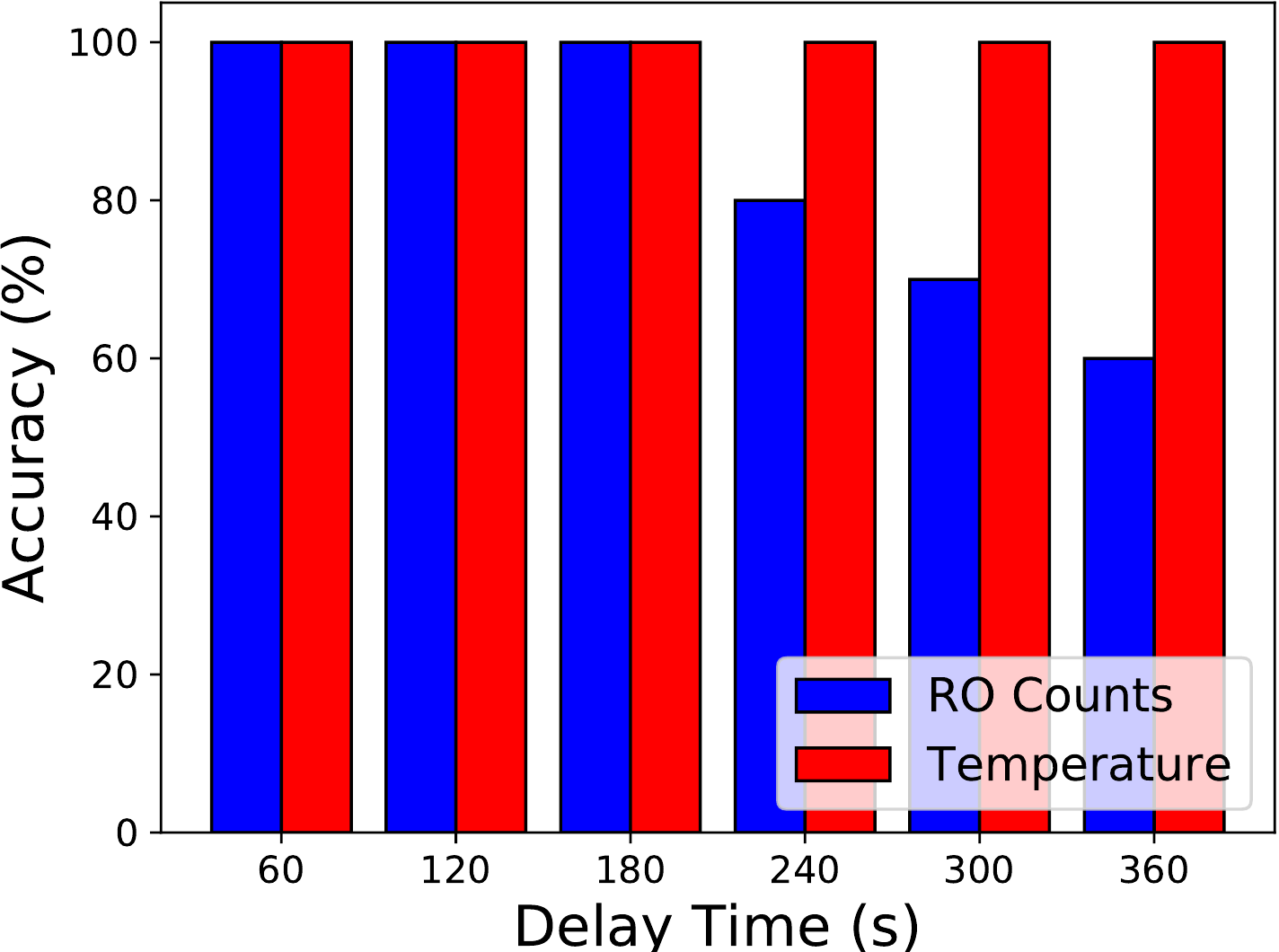}
         \caption{Public cloud server covert channel transmission test accuracy with different delay times.}
         \label{fig:Public server covert channel transmission test accuracy}
     \hfill
\end{figure}

Recall from Section~\ref{sec_covert_channel_design} that the
thermal covert channel uses a simple on-off keying scheme where high temperature, i.e. low RO counts, corresponds to a $1$ and low temperature, i.e. high RO counts, corresponds to a $0$.
To transmit a bit of data, if an $1$ is to be transmitted, the ``sender'' stresses the SmartSSD to achieve the highest temperature. 
If a $0$ is to be transmitted, the sender leaves the SmartSSD to be idle. 
Next, the ``receiver'' user loads his or her FPGA bitstream with the RTL kernel implementing the RO sensor and reads the RO counts. The SmartSSD is then left idle to return to baseline temperature. The receiver can compare the RO counts measured to RO counts corresponding to baseline temperature to determine if the transmitted bit was either $1$ or $0$.

\begin{figure}[t]
     \centering
         \centering
         \includegraphics[height=4.5cm]{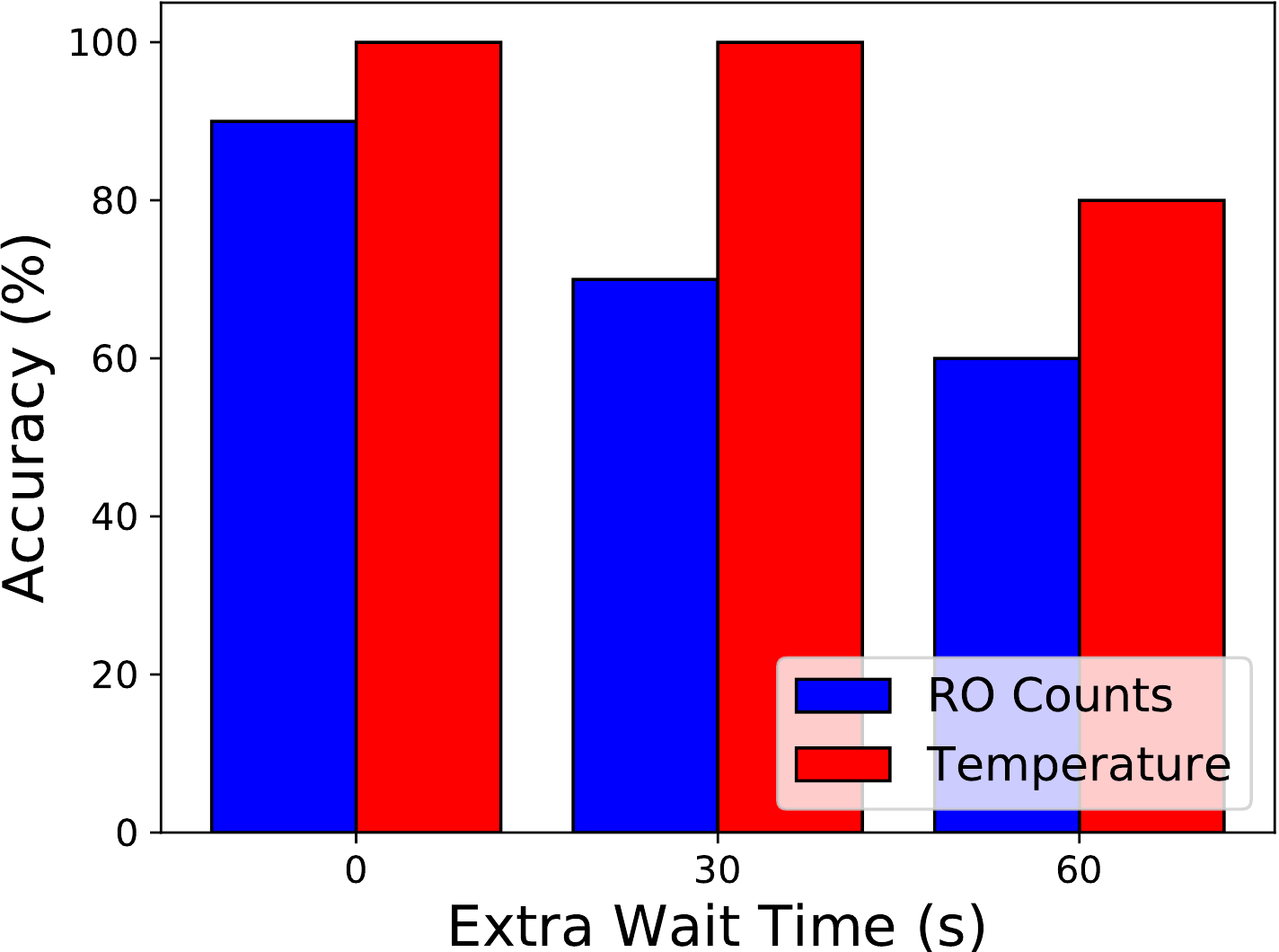}
         \caption{Public cloud server covert channel transmission test accuracy with different extra waiting times for 2 users.}
         \label{fig:Public server covert channel transmission test accuracy between two users}
     \hfill
\end{figure}

For the covert channel evaluation, we randomly select a $1$ or a $0$. We then either stress the SSD or do nothing. We then wait for different amount of delay time, we tested delays from $60$ to $360$ seconds. After the delay,
the SSD temperature measurements as well as the RO count measurements are then taken after waiting $15$ minutes to ensure that the SmartSSD returns to the baseline temperature before proceed to the next bit transmission. To recover the transmitted bit, we set a threshold, which is the RO count's value. The thresholds are set by measuring baseline temperature or RO counts, i.e. when the SSD is idle and then adding a fixed delta to the baseline. 

To evaluate effectiveness of the RO counts, we also measured temperature using the SSD thermal sensors. With SSD thermal sensors, a similar approach is taken of comparing the measured temperature to the baseline temperature.

\subsection{Covert Channel Results}

Our results for the university server and the public cloud server are shown in Figure~\ref{fig:University remote server covert channel transmission test accuracy} and Figure~\ref{fig:Public server covert channel transmission test accuracy}, respectively. It can be clearly seen that we achieve the highest accuracy within the first $4$ to $5$ minutes when using both the SmartSSD temperature and RO counts data. Therefore, within these $4$ to $5$ minutes the heat generated by one user can be observed by another user who later uses the same SmartSSD and as a result a transfer of data through a covert channel takes place. As the delay time increases, it should be noted that the accuracy drops. This is consistent as it becomes difficult to differentiate between a $1$ or $0$ for longer periods of time as the SmartSSD returns to the baseline temperature. It can be observed that after the first $3$ minutes, the accuracy on the public cloud server is significantly lower than the accuracy on the university server. The main reason for this is the more capable cooling system that the public cloud server is equipped with. Given that, on the public cloud server it is  more difficult to differentiate between a $1$ and a $0$ as time passes, although the accuracy is maintained in high standards even after $6$ minutes after stressing.
However, even though the accuracy drops for delays longer than $5$ minutes, transmission is possible and error-correction codes could be used.

For comparison, the accuracy of the covert channel using the SSD thermal sensor is also shown. It can be seen that RO count based covert channel has only about $10$\% lower accuracy. Thus, even if there is no access to the SSD thermal sensors, the attackers can always use the RO sensor based covert communication with high accuracy. Further, Manchester encoding could be used for even better accuracy and thus our evaluation gives conservative results for the accuracy of the novel thermal temporal channel.

\begin{table}[!t]
\caption{Bandwidth for covert channel using one SmartSSD in one and two users scenarios. Total transmission time is calculated as the sum of SSD heat up time, VM instance setup time, bitstream load time and thermal measurement time. SSD heat up time is constant to $300$ seconds, VM instance setup time was tested from $60$ to $360$ seconds in increments of $60$, bit stream load time is measured to be $5$ seconds and thermal measurement time is constant to $60$ seconds. In case of two users scenario, a fixed time of $35$ seconds (time required between two users to switch) is added to the total transmission time.}
\label{table_bandwidth}
\small
\begin{tabular}{|l|l|l|}
\hline 
\textbf{Total Transmission} & \textbf{1 User 1 SSD} & \textbf{2 Users 1 SSD} \\ 
\textbf{Time (s)} & \textbf{Bandwidth (bit/s)} & \textbf{Bandwidth (bit/s)} \\ \hline \hline
425                             & 0.002353         &        0.002174  \\ \hline 
485                              & 0.002062          & 0.001923         \\ \hline
545                             & 0.001835          & 0.001725         \\ \hline
605                              & 0.001653     & 0.001563          \\ \hline
665                              & 0.001504          & 0.001429         \\ \hline
725                              & 0.001379          & 0.001316        \\ \hline
\end{tabular}
\end{table}
We also tested the covert channel between two users.
In Table~\ref{table_bandwidth} we show the transmission bandwidth for one and two users deploying one SmartSSD. While it is understood that the covert channel is slow, users can rent easily more SmartSSDs in parallel to increase the bandwidth of the covert transmission. 
It can be observed that the bandwidth for two users is slightly lower than the bandwidth for one user. The reason for this is that it takes some extra time to set up the experiment once the instance is obtained and there is a switch from one user to another within the cloud scenario. In our experiments, the time required between two users to switch has been measured to be $35$ seconds on average.

We also experimented with adding additional delay in the two-users scenario to test how much it affects the accuracy. In this case, we add more delay, extra from the time required for the second user to set up the newly obtained instance and load the bitstream that is always required. 
Figure~\ref{fig:Public server covert channel transmission test accuracy between two users} shows the covert channel transmission accuracy between two users in a public cloud server. It can be clearly seen that as we increase the extra wait time, the accuracy in both RO counts and temperature drops. As mentioned in Section~\ref{sec_covert_channel_design}, we can refer to Figure~\ref{fig:timeline} to understand more clearly each delay time and how the process goes, as the timeline in switching between two users is shown. The instance is initially dedicated to Alice, where she heats up the SSD. Then, Alice releases the SmartSSD and Bob is able to occupy it. There is some amount of time that Bob requires to wait each time before he is able to measure and get the RO counts. After the instance is dedicated to Bob, Bob needs some amount of time to set up the instance. To set up the instance, we tested from $60$ to $360$ seconds in increments of $60$. Secondly, Bob spends time to load the bitstream. In our experiments, this time has been measured to be $5$ seconds on average. Finally, we tested the accuracy of the proposed covert channel by adding extra wait time after loading the bitstream and before taking the measurements. This time corresponds to $30$, $60$ seconds and no additional waiting time, as we show in Figure~\ref{fig:Public server covert channel transmission test accuracy between two users}.

\section{Related Work}
\label{sec_related_work}

Our work is the first work on security analysis of SmartSSDs. There is, however, relevant existing work, mostly concerning FPGAs. It is now well-known that it is possible to implement temperature sensors suitable for thermal monitoring on FPGAs using ring oscillators~\cite{boemo_thermal_1997}, whose frequency drifts in response to temperature variations~\cite{valtchanov_modeling_2008, lopez_dynamically_2002, lopez_thermal_2000,yin_temperature_2009}. 
For example, existing work~\cite{tian_temporal_2019} has explored a type of temporal thermal attack where heat generated on an FPGA by one ring oscillator heater circuit can be later observed by a ring oscillator sensor circuit that is loaded onto the same FPGA. Our work meanwhile explores how to heat up SSD, the heat retention of the SSD, and how ROs can be used to measure thermal state of the SSD in the same SmartSSD package as the FPGA instantiated with the RO sensors.
This type of attack is able to leak information between different users of an FPGA who are assigned to the same FPGA over time. Our work on SmartSSD follows a similar idea, but is unique to the SSD disk setting.

\section{Conclusion and Future Work}
\label{sec_conclusion}

This work explored security threats to FPGA-enabled SmartSSDs. A SmartSSD is a solid-state disk augmented with an FPGA. The disk and FPGA share a PCIe connection to the host computer and are enclosed in a single package. The purpose of the FPGA is to enable computation on the data stored on the disk, without use of the main host computer. Through public cloud providers, it is now possible to rent SmartSSDs on-demand. The SmartSSDs can be shared by different users, where one user accesses the disk at a time, and then the disk is allocated to another user. The sharing can enable better utilization of the disks, but also leads to new security attacks. This paper in particular showed that heat generated by one user can be observed by another user who later uses the same SmartSSD. Based on the thermal state of the SmartSSD, a covert data transfer can be achieved through simple On-Off Keying (OOK). Use of multiple SmartSSDs in parallel can further, significantly improve the data throughput of the covert channel. The new temporal thermal covert channel in SmartSSDs was demonstrated on a public cloud provider as well as on an in-lab server.

\bibliographystyle{ACM-Reference-Format}
\bibliography{bibtex/references}

\end{document}